\documentclass[twocolumn]{jpsj2} 

\title{Disorder-Driven Quantum Phase Transition from Antiferromagnetic Metal to Insulating State in Multilayered High-$T_{\rm c}$ Cuprate (Cu,C)Ba$_2$Ca$_4$Cu$_5$O$_y$ }
\author{Hidekazu Mukuda$^{1}$\thanks{E-mail address: mukuda@mp.es.osaka-u.ac.jp}, Machiko Abe$^{1}$, Sunao Shimizu$^{1}$, Yoshio Kitaoka$^{1}$, \\Akira Iyo$^{2}$, Yasuharu Kodama$^{2}$, Hijiri Kito$^{2}$, Yasutomo Tanaka$^{2}$, Kazuyasu Tokiwa$^{3}$ and Tsuneo Watanabe$^{3}$}

\inst{$^{1}$Department of Materials Engineering Science, Osaka University, Toyonaka, Osaka 560-8531 \\
$^2$National Institute of Advanced Industrial Science and Technology (AIST), Umezono, Tsukuba 305-8568 \\
$^3$Department of Applied Electronics, Tokyo University of Science, Noda, Chiba 278-8510}

\abst{
We report on superconducting (SC) characteristics for the oxygen-reduced Cu-based five-layered high-temperature superconductor  (Cu,C)Ba$_2$Ca$_4$Cu$_5$O$_y$(Cu-1245(OPT)), which includes pyramidal outer planes (OPs) and  square inner planes (IPs).
As a result of a reduction in the carrier density, the superconductivity for Cu-1245(OPT) occurs at the nearly optimally doped OPs with $T_{\rm c}$= 98 K. The onset of static antiferromagnetic (AFM) order at IPs is evidenced from the observation of zero-field Cu NMR at low temperatures, although the SC transition at OPs emerges below $T_{\rm c}$=98 K. 
A disorder, which is actually mapped onto the underdoped IPs, is demonstrated to cause a quantum phase transition from AFM metal to insulating state in an underdoped regime. This finding reinforces that an AFM metallic phase exists between the AFM insulating phase and the SC phase for the ideally flat CuO$_2$ plane without disorder, as reported for the IPs in optimally doped Hg-1245 with the same doping level as in Cu-1245(OPT). }

\kword{multilayered cuprate, antiferromagnetism, superconductivity, disorder, NMR}

\begin{document}
\maketitle

\date{\today}

The possible coexistence of antiferromagnetism and high-$T_{\rm c}$ superconductivity (HTSC) remains as one of the most interesting problems in high-$T_{\rm c}$ cuprates and will allow us to gain insight into the origin of HTSC\cite{Anderson,Inaba,Lee,Miller,Sidis,Lake,Zhang,Demler,TKLee,Ogata}.  
Multilayered high-$T_c$ cuprates, which have more than three CuO$_2$ planes in a unit cell, exhibit unique magnetic and superconducting (SC) properties.
They include two types of CuO$_2$ planes, an outer CuO$_2$ plane (OP) in a five-fold pyramidal coordination and an inner plane (IP) in a four-fold square coordination.
The unusual magnetic and SC characteristics in Hg- and Cu-based multilayered high-$T_c$ cuprates have been studied through $^{63}$Cu-NMR measurements \cite{Tokunaga,Kotegawa2001,Kotegawa2004,Mukuda}; 
(i) The ideal flatness of CuO$_2$ planes with homogeneously doped carriers, which is ensured by the very narrow Cu-NMR linewidth, is one of the key features that distinguish the multilayered cuprates from other cuprates like La$_{2-x}$Sr$_{x}$CuO$_4$(LSCO). 
(ii) The doping level at OP $N_h$(OP) is larger than $N_h$(IP) at the IP for all systems\cite{Tokunaga} and its difference $\Delta N_h=N_h$(OP)-$N_h$(IP) increases as either the total carrier content or the number of CuO$_2$ planes increases\cite{Kotegawa2001}.
(iii) In five-layered optimally doped HgBa$_2$Ca$_4$Cu$_5$O$_{y}$(Hg-1245(OPT)), the two optimally doped OPs are predominantly superconducting below $T_{\rm c}$ = 108 K, whereas the three underdoped IPs show antiferromagnetic (AFM) metallic behavior below $T_{\rm N}$ = 60 K, realizing the coexistence of an AFM phase and a SC phase in a unit cell\cite{Kotegawa2004}. 
(iv) The first microscopic evidence for the uniform mixing of both phases on a single CuO$_2$ plane has been reported in very underdope Hg-1245(UD) with $T_{\rm N}$ = 290 K and $T_{\rm c}$ = 72 K\cite{Mukuda}. 
The superconductivity with a high $T_{\rm c}$ value surprisingly survives with the metallic AFM order in the multilayered cuprates, and has stimulated theoretical works on the intimate interplay between both phases \cite{Mori}.

In the previous work on an as-prepared sample of (Cu,C)Ba$_2$Ca$_4$Cu$_5$O$_{y}$(Cu-1245(OVD)) with $T_c=90$ K, the Cu-NMR study revealed that the optimally doped IP ($N_h($IP$)\approx0.17$) is responsible for the onset of superconductivity at 90 K, whereas the bulk superconductivity anomaly does not emerge at the overdoped OP($N_h($OP$)\approx0.28$) until 65 K, indicating that the IPs play a primary role in the onset of superconductivity at 90 K \cite{Kotegawa2001}. 
The AFM order has never been observed in the as-prepared Cu-1245(OVD) because it tends to be heavily overdoped\cite{Tokunaga,Kotegawa2001}. By contrast, a static AFM order was evidenced in the as-prepared samples of Hg-1245 and Tl-1245, which are optimally doped and slightly overdoped, respectively\cite{Kotegawa2004,Mukuda}. 
Thus, Cu-1245 systems provide a good opportunity to study the phase diagram of a flat CuO$_2$ plane from the underdoped region to overdoped region.

In this letter, we report a Cu-NMR study on oxygen-reduced Cu-1245(OPT), where the carrier density at both OP and IP is smaller than in the as-prepared Cu-1245(OVD).
We demonstrate that a disorder effect makes carriers localize and eventually drives a quantum phase transition from an AFM metal to an insulating state.

\begin{figure}[h]
\begin{center}
\includegraphics[width=0.8\linewidth]{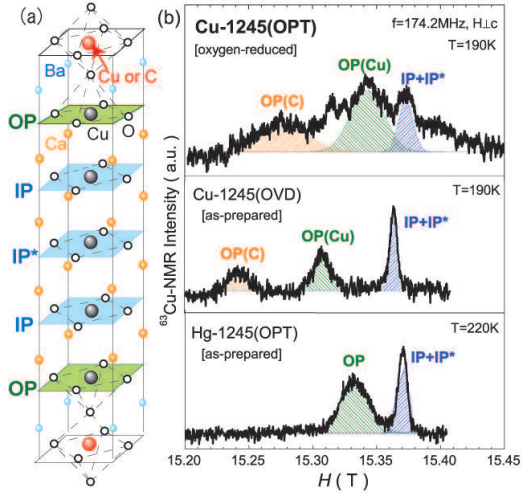}
\end{center}
\caption{(color online) (a) Crystal structure of Cu-1245. (b) Typical $^{63}$Cu-NMR spectra of Cu-1245(OPT) together with those of the as-prepared samples of Cu-1245(OVD) and Hg-1245(OPT). In Cu-1245 systems, there are two Cu sites at OP, since the Cu at the charge reservoir layer is partially substituted by carbon.}
\label{fig:structures}
\end{figure}


Polycrystalline samples of Cu-1245 were prepared by high-pressure synthesis\cite{Ihara,Hirai}. 
To reduce the carrier density in Cu-1245(OVD), the sample was annealed in N$_2$ gas atmosphere at 525$^\circ$C for 12 h. 
$T_{\rm c}$=98 K for deoxidized Cu-1245(OPT) is determined by the onset of a sharp diamagnetic response observed by dc-susceptibility measurement, which is higher than  $T_{\rm c}=$90 K for Cu-1245(OVD). 
From X-ray diffraction, the powdered sample was confirmed to consist of almost a single phase, whereas dc susceptibility indicated that a small fraction of higher-$T_c$ phase is included ($\sim$5\%). 
The $^{63}$Cu-NMR measurement was performed at 174.2 MHz in the field perpendicular to the $c$-axis and in a $T$ range of 10 - 300 K.
 
As shown in Fig. \ref{fig:structures}(b), the $^{63}$Cu-NMR spectra for IPs and OPs are separately observed for Cu-1245(OPT) as well as for other multilayered cuprates \cite{Tokunaga,Kotegawa2001,Kotegawa2004}. 
In particular, in Cu-1245 systems, there are two Cu sites at OP, denoted by OP(C) and OP(Cu), since the Cu at the charge reservoir layer is partially substituted by carbon. 
The NMR intensity ratio of OP(C) to OP(Cu) (see the middle spectrum of Fig. \ref{fig:structures}(b)) suggests that approximately 40\% of the Cu at the charge reservoir layer is substituted by carbon. 
The three peaks for the middle NMR spectrum in Fig. \ref{fig:structures}(b) are distinguished by the difference in electric quadrupole frequency $\nu_Q$: $^{63}\nu_Q$(IP), $^{63}\nu_Q$(OP(Cu)) and $^{63}\nu_Q$(OP(C)) are estimated to be 9.5, 13.7 and 29.2 MHz in Cu-1245(OVD), respectively, by analyzing the field dependence of the NMR spectra. 
Each NMR spectral width in Cu-1245(OPT) is approximately 100 Oe for the IP, 260 Oe for OP(Cu) and 380 Oe for OP(C) at high temperatures, which are about two times larger than those of the as-prepared samples.
This result demonstrates that the disordered potential due to the deficiency of oxygen at the charge reservoir layers is actually mapped  onto the IPs in deoxidized Cu-1245(OVD). 

\begin{figure}[htbp]
\begin{center}
\includegraphics[width=0.75\linewidth]{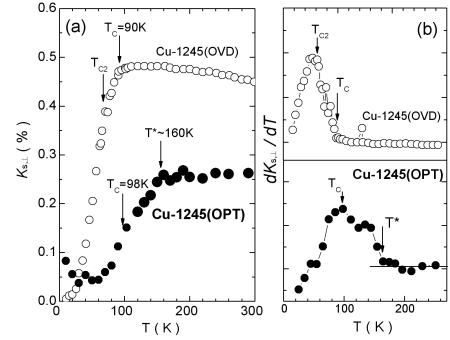}
\end{center}
\caption{(a) Temperature dependence of $K_{\rm s,\perp}$ of the OP at the field perpendicular to the $c$-axis for  deoxidized Cu-1245(OPT) and as-prepared Cu-1245(OVD)\cite{Kotegawa2001}. (b) $T$ dependence of $dK_{\rm s,\perp}$(OP)$/dT$ for both samples. $T_c$ at each CuO$_2$ plane was determined from the maximum in $dK_{\rm s,\perp}$(OP)$/dT$ {\it vs} $T$ plots. }
\label{fig:Knightshift}
\end{figure}

Knight shift was determined from the central peak ($-1/2\Leftrightarrow+1/2$ transition) in the $^{63}$Cu-NMR spectrum. 
The Knight shift consists of the orbital part ($K_{\rm orb}$) and the spin part ($K_{\rm s}$); $K_\alpha= K_{\rm s,\alpha}+K_{\rm orb,\alpha} (\alpha=\perp,\parallel)$. 
Figure \ref{fig:Knightshift}(a) shows the temperature ($T$) dependence of $K_{\rm s,\perp}(T)$ of the OP for $H\perp c$, together with that of Cu-1245(OVD). 
Here we assumed $K_{\rm orb,\perp}$(OP) to be 0.21\% using the value for Cu-1245(OVD). 
$K_s$ at the OP decreases below 160 K, which is higher than $T_c=98$ K, suggesting the existence of a pseudogap below $T^*\sim$ 160 K. 
The local carrier density at OP ($N_h$(OP)) is estimated using the systematic experimental relation between $N_h$ and $K_{\rm s,\perp}$ at room temperature\cite{Kotegawa2001}. 
$N_h$(OP) is estimated to be $\sim$0.18, which indicates that the OP is optimally doped. 
For the IP, $K_s$ at IP cannot be precisely determined owing to the disappearance of NMR spectra below $T\sim$170 K associated with the development of AFM correlation. 
The increase in $K_{\rm s,\perp}$(OP) below $\sim$60 K may be relevant to the onset of AFM order at the IP as well as for Hg-1245(OPT)\cite{Kotegawa2004}. 
It is evident that the carrier density at both the IP and the OP is successfully decreased by reducing oxygen via annealing in N$_2$ gas atmosphere. 
$T_c$ at each CuO$_2$ plane was determined from the maximum of the derivatives of $K_{s,\perp}$ against $T$, $dK_{s,\perp}/dT$. 
Figure \ref{fig:Knightshift}(b) shows the $T$ dependence of $dK_{s,\perp}/dT$, together with that of Cu-1245(OVD). 
It has a maximum at approximately $T_c$=98 K. 
This indicates that the nearly optimally doped OP is responsible for the bulk superconductivity, By contrast, the superconductivity at the overdoped OP for Cu-1245(OVD) is parasitic down to 65 K under the onset of bulk superconductivity below 90 K at the nearly optimally doped IPs.
This gives evidence that optimally doping carriers into either the OP or the IP determines its bulk $T_{\rm c}$ in the multilayered cuprates.

\begin{figure}[htbp]
\begin{center}
\includegraphics[width=0.8\linewidth]{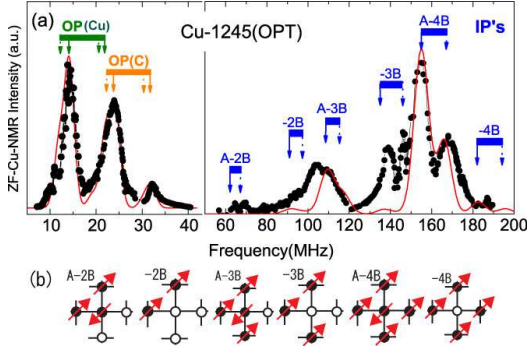}
\end{center}
\caption{(color online) (a) Cu-NMR spectra at 1.5K and zero magnetic field for Cu-1245(OPT). Each discrete peak above 50 MHz arises from IPs, which is qualitatively explained by a carrier-localization model. The red curve is a simulated spectrum for the case that the carrier of $N_h\sim$6\% is distributed randomly. Each peak below 50 MHz arises from OP(Cu) and OP(C). The sets of solid and broken arrows correspond to $^{63}$Cu- and $^{65}$Cu-NMR spectra, respectively. (b) Illustration of possible distribution patterns of localized carriers at low temperatures.  }
\label{fig:ZF-NMR}
\end{figure}

\begin{figure}[htbp]
\begin{center}
\includegraphics[width=0.8\linewidth]{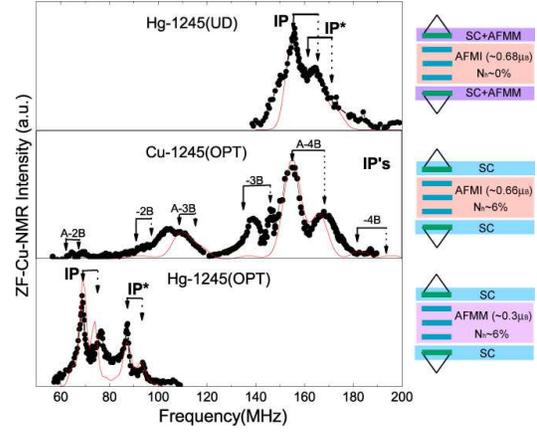}
\end{center}
\caption{(color online) Comparison of ZF-Cu-NMR spectra for IPs at 1.5 K in Cu-1245(OPT) with those in Hg-1245(OPT)\cite{Kotegawa2004} and Hg-1245(UD)\cite{Mukuda}. Right figures schematically indicate the state at each layer for three compounds.}
\label{fig:Cu-Hg}
\end{figure}

The onset of AFM order is suggested from the disappearance of NMR spectra at the IP below $T\sim$170 K as well as for Hg-1245(OPT). 
This is because the carrier densities for both are comparable to one another.
The observation of NMR spectra at zero magnetic field enables us to determine the magnetic ground state, since magnetically ordered moments induce a large internal field ($H_{\rm int}$) at nuclear sites.
Figure \ref{fig:ZF-NMR}(a) shows zero-field (ZF) Cu-NMR spectra at 1.5 K for Cu-1245(OPT). 
Unexpectedly, they consist of discrete resonance peaks over a wide frequency range. 
The spectra above and below 50 MHz are identified as arising from the Cu sites at the IP and the OP, respectively. 
They are markedly different from those of the IPs in optimally doped Hg-1245(OPT) and underdoped Hg-1245(UD), which are shown in the bottom and the top of Fig. \ref{fig:Cu-Hg}, respectively. 
Note that the carrier density at the IP in Cu-1245(OPT) is comparable to that in Hg-1245(OPT). 
We explain the discrete NMR spectra of the IPs in terms of a carrier-localization model, which is described by the discrete values of $H_{\rm int}$.  
These observed resonance frequencies are given by $f_{\rm res}^{(i)}=(\gamma_N/2\pi) H_{\rm int}^{(i)}=(\gamma_N/2\pi) |A_{\rm hf}^{(i)}|M_{\rm AF}$, where $\gamma_N$ is the nuclear gyromagnetic ratio, $A_{\rm hf}^{(i)}$ is the hyperfine coupling constant, and $M_{\rm AF}$ is the AFM moment.
We consider  several possible distribution patterns of the localized carriers, which are shown in Fig. \ref{fig:ZF-NMR}(b). 
Here internal fields are calculated as follows: $(-4B)M_{\rm AF}$, $(A-4B)M_{\rm AF}$, $(-3B)M_{\rm AF}$, $(A-3B)M_{\rm AF}$, $(-2B)M_{\rm AF}$, and $(A-2B)M_{\rm AF}$.
$A$ and $B$ are the on-site hyperfine field and the supertransferred hyperfine field from the four nearest-neighboring Cu sites, respectively, and are assumed to be $A\sim$ 37kOe/$\mu_{\rm B}$ and $B\sim$ 61kOe/$\mu_{\rm B}$, which are the same values as those in Hg-1245 systems \cite{Mukuda}.
$M_{\rm AF}$ is assumed to be 0.66$\mu_{\rm B}$, which is almost the same as that in a nearly nondoped CuO$_2$ plane\cite{Mukuda}.
Actually, each resonance peak is qualitatively reproduced by this simple model, as indicated by the arrows in Fig. \ref{fig:ZF-NMR}(a). 
The curve in the figure is a simulated spectrum for the case that doped holes with $N_h\sim$6\% are distributed randomly, assuming the same carrier density as that in Hg-1245(OPT). 
Note that the error in $N_h$ is less than 2\% when taking account of the ambiguity of the intensity, particularly for the case of widely distributed NMR spectra. 
The NMR intensities are larger than those shown by the simulated curve for the resonances with $H_{\rm int}^{(i)}=(-3B)M_{\rm AF}$ and $(-2B)M_{\rm AF}$, where two or more holes are gathered. 
This fact implies that two or more holes tend to become attractive at low temperatures even though carrier density is very low. 
In addition, each resonance peak in this spectrum is significantly broader than for the others, as shown in Fig. \ref{fig:Cu-Hg}.
This suggests that the carriers are holes doped on the oxygen $2p$ states, which hybridize with the nearest Cu-$3d_{x^2-y^2}$ state. 
In this case, if holes at oxygen sites are localized, their surrounding Cu spins are reduced owing to the $d$-$p$ hybridization, resulting in the NMR spectrum of Cu-1245(OPT) being broader than for the Hg-1245(OPT) and Hg-1245(UD). 

\begin{figure}[htbp]
\begin{center}
\includegraphics[width=0.8\linewidth]{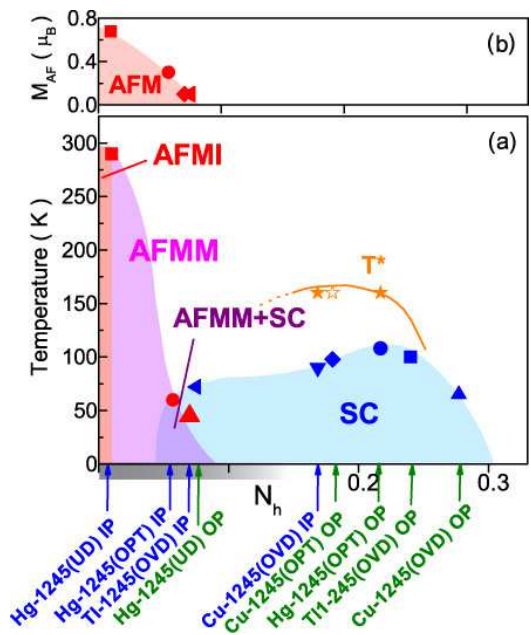}
\end{center}
\caption{(color online)(a) Phase diagram for various five-layered HTSCs. Here we denote the AFM insulator and metal phases as AFMI and AFMM, respectively. $T^*$ is the pseudogap temperature deduced from a decrease in $1/T_1T$ (solid stars) and $K_s$ (open star). Panel (b) shows the variation in size of the AFM ordered moment $M_{\rm AF}$ as a function of hole density. This phase diagram is revised from the previous report\cite{Mukuda}.}
\label{fig:Phasediagram}
\end{figure}

The magnetic state at the IPs in Cu-1245(OPT) is an AFM insulating state, which contrasts with the AFM metal with reduced AFM moments 0.3-0.37$\mu_{\rm B}$ in Hg-1245(OPT) \cite{Mukuda}. 
This difference comes from the disorder effect mapped on the IPs due to the deficient oxygen at the charge reservoir layers in Cu-1245(OPT), as corroborated by the comparison of the Cu-NMR spectral width with those of as-prepared samples, which are shown in Fig. \ref{fig:structures}(b). 
As a result, the carriers are localized owing to the Anderson-localization mechanism, since a small amount of carriers are doped into two-dimensional IPs near the metal-insulator transition. 
The carrier localization in the AFM metallic state has also been observed in the four-layered cuprates Ba$_2$Ca$_3$Cu$_4$O$_8$(O$_{1-y}$F$_{y}$)$_2$, where the apical oxygen (O$^{2-}$) at the OP is partially substituted by fluorine (F$^-$)\cite{Shimizu}. 
In this context, a reason why the IPs for Hg-1245(OPT) with $T_c=108$ K and $T_N=60$ K maintains the uniformly reduced AFM moment of 0.3-0.37$\mu_B$ is that the AFM metallic state is realized down to 1.5 K. 
In Cu-1245(OPT) with a nearly equal doping level $N_h\sim 6$ \% to that in Hg-1245(OPT), however, the disorder mapped on the IPs from either the charge reservoir layers or the OP makes carriers localize at low temperatures, causing the disorder-driven quantum phase transition from an AFM metal to an insulating state. 
We note furthermore that in the case of oxygen-reduced Hg-1245(UD)\cite{Mukuda}, a thorough re-examination of the ZF Cu-NMR spectra at 1.5 K revealed that there were no other resonance peaks except for ones arising from the Cu sites with $(A-4B)M_{\rm AF}$, as shown in Fig. \ref{fig:Cu-Hg}. This ensures that the IPs in Hg-1245(UD) with $T_c=72$ K and $T_N$=290 K are almost nondoped. 
Consequently, these findings reinforce the fact that the AFM metallic phase exists between the AFM insulating state  and the SC state for an ideally flat CuO$_2$ plane without disorder (see Fig. \ref{fig:Phasediagram}). 
Furthermore, since the disorder effect is inevitable because of the chemical substitution introduced into the out-of-planes in LSCO, it is likely that the spin-glass phase emerging in the range $0.02 < N_h <0.05$ is caused by disorder.

For the OP for Cu-1245(OPT), the Cu-NMR spectra at zero field for OP(Cu) and OP(C) are split into two segregated peaks around respective $^{63}\nu_Q$(OP(Cu)) and $^{63}\nu_Q$(OP(C)), respectively owing to the presence of a small $H_{\rm int}\sim$0.34 T for both, as shown in Fig. \ref{fig:ZF-NMR}(a). 
This is comparable to $H_{\rm int}\sim0.54$ T for the OP of Hg-1245(OPT), which has been explained by the proximity effect of the AFM-ordered IPs\cite{Kotegawa2004}. 
$T_c$(OP) for Cu-1245 (OPT) is plotted in the phase diagram of Fig. \ref{fig:Phasediagram} as a function of local carrier density at a CuO$_2$ plane together with $T_c$(OP, IP) for Cu-1245(OVD). 

Finally, we deal with the pseudogap behavior in the multilayered cuprates. 
A pseudogap temperature $T^*$, below which $1/T_1T$ starts to decrease, is plotted in Fig. \ref{fig:Phasediagram}(a) for the OP in Hg-1245(OPT) and the IP in Cu-1245(OVD)\cite{Kotegawa2004}. 
$K_s$ at the OP in Cu-1245(OPT) decreases below 160 K, suggesting the existence of a pseudogap below $T^*\sim$ 160 K. 
In the overdoped region, a pseudogap is not observed for the OP in Tl-1245(OVD).
On the other hand, for the heavily underdoped IPs in Hg-1245(OPT) and Tl-1245(OVD), $1/T_1T$ does not exhibit any sign of pseudogap behavior \cite{Kotegawa2004}, but instead, the development of low-energy spin excitations prevents us from observing the Cu-NMR spectra, and as a result, the AFM order takes place.  
In this context, the pseudogap behavior observed in the NMR experiments may be intimately related to  superconducting fluctuations where SC-phase coherence remains finite\cite{Ong}, which differs from the pseudogap behavior observed in the one-particle spectral weight probed by ARPES\cite{T*ARPES1,T*ARPES2}.

Through the systematic studies on the multilayered cuprates, one of the surprises is why Cu-1245(OPT) maintains a high $T_{\rm c}$ value of 98 K, even though the AFM insulating IPs  are inserted into the superconducting OPs. 
$T_{\rm c}=$ 98 K  for Cu-1245(OPT) is higher than $T_{\rm c}=$ 90 K  for Cu-1245(OVD) after the deoxidization of Cu-1245(OVD) whose IPs and OPs are both superconducting. 
This implies that  $T_{\rm c}$ is determined by a doping level at a single layer, regardless of what occurs in the nearest-neighboring planes, which are distant in atomic scales. 
By coincidence, $T_{\rm c}$=98 K for Cu-1245(OPT) is the same as that for single-layered Hg-1201 \cite{Hg1201}, suggesting that a single optimally doped flat CuO$_2$ plane shows HTSC up to approximately 100 K. 


In conclusion, we have investigated the superconducting characteristics for the oxygen-reduced Cu-based five-layered high-$T_{\rm c}$ cuprates Cu-1245(OPT), and compared them with those for the previously reported five-layered cuprates.
It has been revealed that bulk superconductivity occurs at the nearly optimally doped OP with $T_{\rm c}$=98 K. This contrasts with the superconductivity for the heavily overdoped Cu-1245(OVD), which emerges at the optimally doped IP with $T_{\rm c}=90$ K. 
This gives evidence that optimally doping carriers into either the IP or the OP determines its bulk $T_{\rm c}$ in the multilayered cuprates.
The ZF-NMR experiments revealed that the static AFM order at the IPs in Cu-1245(OPT) is caused  by the localization of holes due to the disorder effect, although the SC transition takes place at OPs below $T_{\rm c}$=98 K. 
This is in remarkable contrast with the AFM metallic phase in Hg-1245(OPT) with the same doping level as that in Cu-1245(OPT).
This finding reinforces the fact that the AFM metallic phase exists between the AFM insulating state and the SC state for an ideally flat CuO$_2$ plane without disorder. 

This work was supported by a Grant-in-Aid for Creative Scientific Research (15GS0213) from the Ministry of Education, Culture, Sports, Science and Technology (MEXT) and the 21st Century COE Program (G18) by the Japan Society of the Promotion of Science (JSPS). 


\end{document}